\documentclass[dvips,12pt,a4paper]{article}
\usepackage{epsfig}
\usepackage{a4wide}
\begin{document}
\pagestyle{empty}
\begin{titlepage}
%\rightline{}
\vspace{1. truecm}
\begin{center}
\begin{Large}
{\bf Search for New Physics Effects at a Linear Collider with Polarized Beams
\footnote
{Talk given at the International School-Seminar "The Actual 
Problems of Particle Physics", Gomel,  August 7 -- 16, 2001
}
}
\end{Large}

\vspace{2.0cm}

{\large  A. Babich and A. Pankov}
\\[0.3cm]

The Pavel Sukhoi Gomel State Technical University, 246746 Belarus

\end{center}
\vspace{1.5cm}

\begin{abstract}
\noindent
The four-fermion contact-interaction searches in the process
$e^+e^-\to\mu^+\mu^-$ at a future
$e^+e^-$ Linear Collider with c.m. energy $\sqrt{s}=0.5$ TeV and
with both beams longitudinally polarized are studied.  
We evaluate the corresponding model-independent constraints 
on the coupling constants, emphasizing the role of beam polarization, 
and make a comparison with the case of Bhabha scattering. 

\noindent

\vspace*{3.0mm}

\noindent
\end{abstract}
\end{titlepage}

\pagestyle{plain}

\section{Introduction}
Contact interaction Lagrangians (CI) generally represent an effective
description of the `low energy' manifestations of some non-standard
dynamics acting at new, intrinsic, mass scales much higher than the
energies reachable at current particle accelerators. As such, they can
be studied through deviations of the experimental observables from the
Standard Model (SM) expectation, that reflect the additional presence of
the above-mentioned new interaction.
Typical examples are the composite models and the exchanges of extremely
heavy neutral gauge bosons and
leptoquarks \cite{Barger:1998nf,Altarelli:1997ce}.
\par
Clearly, such deviations are expected to be extremely small, as they
would be suppressed for dimensional reasons by essentially some power of
the ratio between the available energy and the large mass scales.
Accordingly, very high energy reactions in experiments with high
luminosity are one of the natural tools to investigate signatures of
contact interaction couplings.
In general, these constants are considered as {\it a priori} free
parameters, and one can quantitatively derive an assessment of the
attainable
reach and of the corresponding upper limits, essentially by numerically
comparing the deviations with the expected experimental statistical and
systematical uncertainties on the cross sections.
\par
Here, we consider the muon pair production process
\begin{equation}
e^++e^-\to \mu^+\mu^-, \label{proc}
\end{equation}
%\rightline{proc}
at a Linear Collider (LC) with c.m.\ energy
$\sqrt s=0.5\hskip 2pt{\rm TeV}$ and polarized electron and positron
beams. We discuss the sensitivity of this reaction to the general,
$SU(3)\times SU(2)\times U(1)$
symmetric $eeff$ contact-interaction effective Lagrangian, with
helicity-conserving and flavor-diagonal fermion
currents \cite{Eichten:1983hw}:
\begin{equation}
{\cal L}_{\rm CI}
=\sum_{\alpha\beta}g^2_{\rm eff}\hskip 2pt\epsilon_{\alpha\beta}
\left(\bar e_{\alpha}\gamma_\mu e_{\alpha}\right)
\left(\bar \mu_{\beta}\gamma^\mu \mu_{\beta}\right).
\label{lagra}
\end{equation}
%\rightline{lagra}
In Eq.~(\ref{lagra}): $\alpha,\beta={\rm L,R}$ denote left- or
right-handed helicities, generation and color indices have been suppressed,
and the CI coupling constants are parameterized in terms of corresponding
mass scales as
$\epsilon_{\alpha\beta}={\eta_{\alpha\beta}}/{{\Lambda^2_{\alpha\beta}}}$
with $\eta_{\alpha\beta}=\pm 1,0$ depending on the chiral structure of the
individual interactions. Also, conventionally the value of $g^2_{\rm eff}$
is fixed at $g^2_{\rm eff}=4\pi$, as a reminder that, in the case of
compositeness, the new interaction would become strong at $\sqrt s$ of the
order of $\Lambda_{\alpha\beta}$. Obviously, in this parameterization,
exclusion ranges or upper limits on the CI couplings can be equivalently
expressed as exclusion ranges or lower bounds on the corresponding mass
scales $\Lambda_{\alpha\beta}$.
\par
For a given final lepton flavour $\mu$, ${\cal L}_{\rm CI}$ in
Eq.~(\ref{lagra}) envisages the existence of eight individual,
and independent, CI models corresponding to the combinations of the four
chiralities $\alpha,\beta$ with the $\pm$ signs of the $\eta$'s,
with {a priori} free, and nonvanishing, coefficients. Correspondingly,
the most general (and model-independent) analysis of the process
(\ref{proc}) must account for the complicated situation where all
four-fermion effective couplings defined in Eq.~(\ref{lagra}) are
simultaneously allowed in the expression for the cross section, and in
principle can interfere and weaken the bounds in case of accidental
cancellations.
\par
Of course, the different helicity amplitudes, as such, do not interfere.
However, {\it the deviations from the SM} could be positive for one
helicity amplitude, and negative for another.
Thus, cancellations might occur.
\par
The simplest attitude is to assume non-zero values for only one of the
couplings (or one specific combination of them) at a time, with all others
zero, this leads to tests of the specific models mentioned above.
But, in principle, constraints obtained by simultaneously
including couplings of different chiralities might become considerably
weaker. Therefore, it
should be higly desirable to apply a more general (and model-independent)
approach to the analysis of experimental data, that simultaneously
includes all terms of Eq.~(\ref{lagra}) as independent free parameters,
and can also allow the derivation of separate constraints (or exclusion
regions) on the values of the coupling constants.
\par
To this aim, in the case of the process (\ref{proc}) at the LC
considered here,
a possibility is offered by initial beam polarization, that
enables us to extract from the data the individual helicity cross sections
through the definition of particular, and optimal, polarized integrated
cross sections and, consequently, to disentangle the constraints on
the corresponding CI constants
\cite{Pankov:1998ad}--\cite{Babich:2001ik}.
In this note, we wish to to present a model-independent analysis of the CI
that complements that of Refs.~\cite{Pankov:1998ad}--\cite{Babich:2001ik},
and is based on the measurements of the observables
such as the total cross section, the forward-backward asymmetry $A_{\rm
FB}$, the left-right asymmetry $A_{\rm LR}$, and left-right forward-backward
asymmetry $A_{\rm LR,FB}$.

%%%%%%%%%%%%%%%%%%%%%%%%%%%%%%%%%%%%%%%%%%%%%%%%%%%%%%%%%%%%%%%%%%%%%%%%
\section{Observables}
%%%%%%%%%%%%%%%%%%%%%%%%%%%%%%%%%%%%%%%%%%%%%%%%%%%%%%%%%%%%%%%%%%%%%%%%
For the process Eq.~(\ref{proc}) we can neglect fermion masses with respect
to
$\sqrt s$, and express the amplitude in the Born approximation including
the
$\gamma$ and $Z$ $s$-channel exchanges plus the contact-interaction term of
Eq.~(\ref{lagra}). With $P_e$ and $P_{\bar e}$ the longitudinal
polarizations
of the electron and positron beams, respectively, and $\theta$ the angle
between the incoming electron and the outgoing fermion in the c.m.\ frame,
the differential cross section can be expressed as \cite{Schrempp:1988zy}:
\begin{equation}
\frac{{\rm d}\sigma}{{\rm d}\cos\theta}
=\frac{3}{8}
\left[(1+\cos\theta)^2 {\sigma}_+
+(1-\cos\theta)^2 {\sigma}_-\right].
\label{cross}
\end{equation}
%\rightline{cross}
In terms of the helicity cross sections $\sigma_{\alpha\beta}$ (with
$\alpha,\beta={\rm L,R}$), directly related to the individual CI
couplings $\epsilon_{\alpha\beta}$:
\begin{eqnarray}
{\sigma}_{+}&=&\frac{1}{4}\,
\left[(1-P_e)(1+P_{\bar{e}})\,\sigma_{\rm LL}
+(1+P_e)(1- P_{\bar{e}})\,\sigma_{\rm RR}\right]\nonumber \\
&=&\frac{D}{4}\,\left[(1-P_{\rm eff})\,\sigma_{\rm LL}
+(1+P_{\rm eff})\,\sigma_{\rm RR}\right],
\label{s+} \\
{\sigma}_{-}&=&\frac{1}{4}\,
\left[(1-P_e)(1+ P_{\bar{e}})\,\sigma_{\rm LR}
+(1+P_e)(1-P_{\bar{e}})\,\sigma_{\rm RL}\right] \nonumber \\
&=&
\frac{D}{4}\,\left[(1-P_{\rm eff})\,\sigma_{\rm LR}
+(1+P_{\rm eff})\,\sigma_{\rm RL}\right], \label{s-}
\end{eqnarray}
where
\begin{equation}
P_{\rm eff}=\frac{P_e-P_{\bar{e}}}{1-P_eP_{\bar{e}}}
\label{pg}
\end{equation}
is the effective polarization \cite{Flottmann:1995ga},
$\vert P_{\rm eff}\vert\leq 1$, and
$D=1-P_eP_{\bar{e}}$. For unpolarized positrons
$P_{\rm eff}\rightarrow P_e$ and $D\rightarrow 1$,
but with $P_{\bar{e}}\ne0$, $\vert P_{\rm eff}\vert$
can be larger than $|P_e|$.
Moreover, in Eqs.~(\ref{s+}) and (\ref{s-}):
\begin{equation}
\sigma_{\alpha\beta}=\sigma_{\rm pt}
\vert{\cal M}_{\alpha\beta}\vert^2,
\label{helcross}
\end{equation}
%\rightline{helcross}
where $\sigma_{\rm pt}\equiv\sigma(e^+e^-\to\gamma^\ast\to l^+l^-)
=(4\pi\alpha^2)/(3s)$.
The helicity amplitudes ${\cal M}_{\alpha\beta}$ can be written as
\begin{equation}
{\cal M}_{\alpha\beta}=Q_e Q_f+g_\alpha^e\,g_\beta^{\mu}\,\chi_Z+
\frac{s}{\alpha}\,\epsilon_{\alpha\beta}
\label{amplit}
\end{equation}
%\rightline{amplit}
where $\chi_Z=s/(s-M^2_Z+iM_Z\Gamma_Z)$ represents the $Z$ propagator,
$g_{\rm L}^{\mu}=(-1/2+ s_W^2)/s_W c_W$ and
$g_{\rm R}^{\mu}=s_W^2/s_W c_W$
are the SM left- and right-handed fermion couplings of the $Z$
with $s_W^2=1-c_W^2\equiv \sin^2\theta_W$.
\par
We now define, with $\epsilon$ the experimental efficiency for
detecting the final state under consideration, the four, directly
measurable,
integrated event rates:
\begin{equation}
N_{\rm L,F},\quad  N_{\rm R,F},\quad N_{\rm L,B},\quad N_{\rm R,B},
\label{obsn}
\end{equation}
where ($\alpha={\rm L,R}$)
\begin{equation}
N_{\alpha,{\rm F}}=\frac{1}{2}{\cal L}_{\rm int}\,\epsilon
\int_{0}^{1}({\rm d}\sigma_\alpha/{\rm d}\cos\theta){\rm d}\cos\theta,
\label{nf}
\end{equation}
\begin{equation}
N_{\alpha,{\rm B}}=\frac{1}{2}{\cal L}_{\rm int}\,\epsilon
\int_{-1}^{0}({\rm d}\sigma_\alpha/{\rm d}\cos\theta){\rm d}\cos\theta,
\label{nb}
\end{equation}
and subscripts R and L correspond to two sets of beam polarizations,
$P_e=+P_1$, $P_{\bar e}=-P_2$ ($P_{1,2}>0$) and $P_e=-P_1$,
$P_{\bar e}=+P_2$, respectively, or, alternatively, $P_{\rm eff}=\pm P$
with $D$ fixed. In Eqs.~(\ref{nf}) and (\ref{nb}), ${\cal L}_{\rm int}$ is
the
time-integrated luminosity, we assume it to be equally distributed over
the two combinations of beam polarizations, L and R.
\par
The set of `conventional' observables we consider here for the discussion
of
bounds on the CI parameters are the unpolarized cross section:
\begin{equation}
\label{unpol}
\sigma_{\rm unpol}
=\frac{1}{4}\left[\sigma_{\rm LL}+\sigma_{\rm LR}
+\sigma_{\rm RR}+\sigma_{\rm RL}\right];
\end{equation}
the (unpolarized) forward-backward asymmetry:
\begin{equation}
A_{\rm FB}=
\frac{3}{4}\,
\frac{\sigma_{\rm LL}-\sigma_{\rm LR}+\sigma_{\rm RR}-\sigma_{\rm RL}}
     {\sigma_{\rm LL}+\sigma_{\rm LR}+\sigma_{\rm RR}+\sigma_{\rm RL}};
\label{afbth}
\end{equation}
the left-right and the left-right forward-backward asymmetries
(which both require polarization), that can be written as, respectively:
\begin{equation}
A_{\rm LR}=
\frac{\sigma_{\rm LL}+\sigma_{\rm LR}-\sigma_{\rm RR}-\sigma_{\rm RL}}
     {\sigma_{\rm LL}+\sigma_{\rm LR}+\sigma_{\rm RR}+\sigma_{\rm RL}},
\label{alrth}
\end{equation}
and
\begin{equation}
A_{\rm LR,FB}=
\frac{3}{4}\,
\frac{\sigma_{\rm LL}-\sigma_{\rm RR}+\sigma_{\rm RL}-\sigma_{\rm LR}}
{\sigma_{\rm LL}+\sigma_{\rm RR}+\sigma_{\rm RL}+\sigma_{\rm LR}}.
\label{alrfbth}
\end{equation}
Using Eqs.~(\ref{helcross}) and (\ref{amplit}), one can easily express
the deviations of these observables from the SM predictions in terms of the
SM couplings and the CI couplings $\epsilon_{\alpha\beta}$ of
Eq.~(\ref{lagra}).
\par
The above observables are connected to the measured ones through the
integrated event rates $N_{\rm L,R;F,B}$, see Eq.~(\ref{obsn}),
as follows:
\begin{equation}
D\,\sigma_{\rm unpol}=\frac{N_{\rm tot }^{\rm exp}}{{\cal L}_{\rm
int}\,\epsilon}, \label{ntotth}
\end{equation}
where
\begin{equation}
N_{\rm tot}^{\rm exp}=N_{\rm L,F}+N_{\rm R,F}+N_{\rm L,B}+N_{\rm R,B}
\label{ntot}
\end{equation}
is the total number of events observed with polarized beams
(for the four measurements).
Eq.~(\ref{ntotth}) expresses the well-known fact that, when both
the electron and positron beams are  polarized, the total annihilation
cross section into fermion-antifermion pairs will be increased by
the factor $D$, with $1\leq D\leq 2$.
\par
For the experimental forward-backward asymmetry:
\begin{equation}
A_{\rm FB}=A_{\rm FB}^{\rm exp}\equiv
\frac{N_{\rm L,F}+N_{\rm R,F}-N_{\rm L,B}-N_{\rm R,B}}
{N_{\rm L,F}+N_{\rm R,F}+N_{\rm L,B}+N_{\rm R,B}},
\label{afbexp}
\end{equation}
Finally, for the experimental left-right and left-right forward-backward
asymmetries the relations are
\begin{equation}
P_{\rm eff}A_{\rm LR}=A_{\rm LR}^{\rm exp}\equiv
\frac{N_{\rm L,F}+N_{\rm L,B}-N_{\rm R,F}-N_{\rm R,B}}
           {N_{\rm L,F}+N_{\rm L,B}+N_{\rm R,F}+N_{\rm R,B}},
\label{alrexp}
\end{equation}
and
\begin{equation}
P_{\rm eff}A_{\rm LR,FB}=A_{\rm LR,FB}^{\rm exp}\equiv
\frac{(N_{\rm L,F}-N_{\rm R,F})-(N_{\rm L,B}-N_{\rm R,B})}
            {N_{\rm L,F}+N_{\rm R,F} + N_{\rm L,B}+N_{\rm R,B}}.
\label{alrfbexp}
\end{equation}
\par
In the following analysis, cross sections will be evaluated including
initial- and final-state radiation by means of the program
ZFITTER \cite{zfitter}, which has to be used along with ZEFIT, adapted to
the present discussion, with $m_{\rm top}=175$~GeV and
$m_H=120$~GeV. One-loop SM electroweak corrections are accounted for by
improved Born amplitudes \cite{Consoli:1989pc,Altarelli:1990dt},
such that the forms of the
previous formulae remain the same. Concerning initial-state radiation, a
cut
on the energy of the emitted photon $\Delta=E_\gamma/E_{\rm beam}=0.9$ is
applied for $\sqrt s=0.5\ {\rm TeV}$ in order to avoid the radiative return
to the $Z$ peak, and increase the signal originating from the contact
interaction contribution \cite{Djouadi:1992sx}.
\par
As numerical inputs, we shall assume the commonly used
reference values of the identification efficiencies
\cite{Damerell}:
$\epsilon=95\%$ for $\mu^+\mu^-$. Concerning the statistical
uncertainty, to study the relative roles of statistical and systematic
uncertainties we shall vary ${\cal L}_{\rm int}$ from $50$ to
$500\ \mbox{fb}^{-1}$ (half for each polarization orientation) with
uncertainty $\delta{\cal L}_{\rm int}/{\cal L}_{\rm int}=0.5\%$, and a
fiducial
experimental angular range $|\cos\theta|\le 0.99$. Also, regarding electron
and
positron degrees of polarization, we shall consider the values: $\vert
P_e\vert=0.8$; $\vert P_{\bar e}\vert=0.0,\ 0.4\ {\rm and}\ 0.6$, with
$\delta P_e/P_e=\delta P_{\bar e}/P_{\bar e}=0.5\ \%$.

%%%%%%%%%%%%%%%%%%%%%%%%%%%%%%%%%%%%%%%%%%%%%%%%%%%%%%%%%%%%%%%%%%%%%%%%%%%%%%
\section{Model independent constraints}
%%%%%%%%%%%%%%%%%%%%%%%%%%%%%%%%%%%%%%%%%%%%%%%%%%%%%%%%%%%%%%%%%%%%%%%%%%%%%%
The current bounds on $\Lambda_{\alpha\beta}$ cited in Sect.~1,
of the order of several TeV, are such that for the LC c.m.\ energy
$\sqrt{s}=0.5$~TeV the characteristic suppression factor $s/\Lambda^2$
in Eq.~(\ref{amplit}) is rather strong. Accordingly, we can safely assume
a linear dependence of the cross sections on the parameters
$\epsilon_{\alpha\beta}$. In this regard, indirect manifestations of
the CI interaction (\ref{lagra}) can be looked for, {\it via} deviations
of the measured observables from the SM predictions, caused by the
new interaction.
The reach on the CI couplings, and the corresponding constraints on their
allowed values in the case of no effect observed, can be estimated
by comparing the expression of the mentioned deviations with the expected
experimental (statistical and systematic) uncertainties.
\par
To this purpose, assuming the data to be well described by the SM
($\epsilon_{\alpha\beta}=0$) predictions, i.e., that no deviation is
observed within the foreseen experimental uncertainty, and in the linear
approximation in $\epsilon_{\alpha\beta}$ of the observables
(\ref{unpol})--(\ref{alrfbth}), we apply the method based on the covariance
matrix:
\begin{eqnarray}
\label{covarv}
&&V_{kl}=\langle({\cal O}_k-\bar{\cal O}_k)
                ({\cal O}_l-\bar{\cal O}_l)\rangle \nonumber \\
&&=
\sum_{i=1}^{4}\left(\delta N_i\right)^2
\left(\frac{\partial{\cal O}_k}{\partial N_i}\right)
\left(\frac{\partial{\cal O}_l}{\partial N_i}\right)
+\left(\delta {\cal L}_{\rm int} \right)^2
\left(\frac{\partial{\cal O}_k}{\partial{\cal L}_{\rm int}}\right)
\left(\frac{\partial{\cal O}_l}{\partial{\cal L}_{\rm int}}\right)
\nonumber \\
&&+
\left(\delta P_e\right)^2
\left(\frac{\partial{\cal O}_k}{\partial P_e}\right)
\left(\frac{\partial{\cal O}_l}{\partial P_e}\right)+
\left(\delta P_{\bar e}\right)^2
\left(\frac{\partial{\cal O}_k}{\partial P_{\bar e}}\right)
\left(\frac{\partial{\cal O}_l}{\partial P_{\bar e}}\right).
\end{eqnarray}
Here, the $N_i$ are given by Eq.~(\ref{obsn}), so that the statistical
error appearing on the right-hand-side is given by
\begin{equation}
\delta N_i=\sqrt{N_i}, \label{deltani} \end{equation}
and the ${\cal O}_l=(\sigma_{\rm unpol}$, $A_{\rm FB}$, $A_{\rm LR}$,
$A_{\rm LR,FB})$ are the four observables. The second, third and fourth
terms of the right-hand-side of Eq.~(\ref{covarv}) represent the systematic
errors on the integrated luminosity ${\cal L}_{\rm int}$, polarizations $
P_e$
and $P_{\bar e}$, respectively, for which we assume the numerical values
reported in the previous Section. From the explicit expression of the
matrix
elements $V_{kl}$, one can easily notice that, apart from
$\sigma_{\rm unpol}$ and $A_{\rm FB}$ that are uncorrelated ($V_{12}=0$),
all other pairs of observables show a correlation. Indeed, the non-zero
diagonal entries are given by:
\begin{eqnarray}
\label{Eq:V-diag}
&&V_{11}= \frac{\sigma_{\rm unpol}^2}{N_{\rm tot}^{\rm exp}}
+ \sigma_{\rm unpol}^2 \,
\left[ \frac{P^2_e P^2_{\bar e}}{D^2}(\epsilon^2_e +\epsilon^2_{\bar e})
+ \epsilon^2_{\cal L} \right]; \nonumber \\
&&V_{22}=\frac{1-A^2_{\rm FB}}{N_{\rm tot}^{\rm exp}}, \hskip 4pt
V_{33}=\frac{1-A^2_{\rm LR} P^2_{\rm eff}}{P^2_{\rm eff} N_{\rm tot}^{\rm
exp}}
+ A^2_{\rm LR} \, \Delta^2_2; \nonumber \\
&&V_{44}=\frac{1-A^2_{\rm LR,FB} P^2_{\rm eff}}
{P^2_{\rm eff} N_{\rm tot}^{\rm exp}} + A^2_{\rm LR,FB}\, \Delta^2_2,
\label{diagv}
\end{eqnarray}
and, for the non-diagonal ones we have:
\begin{eqnarray}
&&V_{13}= \sigma_{\rm unpol} \, A_{\rm LR} \, \Delta^2_1; \qquad
V_{14}= \sigma_{\rm unpol}\,
A_{\rm LR, FB} \, \Delta^2_1; \nonumber \\
&&V_{23}=\frac{A_{\rm LR,FB}-A_{\rm FB} A_{\rm LR}}{N_{\rm tot}^{\exp}},
\hskip 4pt
V_{24}=\frac{A_{\rm LR}-A_{\rm FB} A_{\rm LR,FB}}{N_{\rm tot}^{\rm exp}};
\nonumber \\
&&V_{34}=\frac{A_{\rm FB}-A_{\rm LR} A_{\rm LR,FB}
P^2_{\rm eff}}{P^2_{\rm eff} N_{\rm tot}^{\rm exp}} + A_{\rm LR}\,
A_{\rm LR,FB}\, \Delta^2_2.
\label{nondiagv}
\end{eqnarray}
Here:
\begin{eqnarray}
&&\Delta^2_1=\frac{P_e \, P_{\bar e}}{P_{\rm eff} D^3}
\left[-(1-P^2_{\bar e})\, P_e \, \epsilon^2_e + (1-P^2_e)\,
P_{\bar e} \, \epsilon^2_{\bar e} \right],  \nonumber \\
&&\Delta^2_2=\frac{(1-P^2_{\bar e})^2 \, P^2_e \, \epsilon^2_e
+ (1-P^2_e)^2 \, P^2_{\bar e} \, \epsilon^2_{\bar e}}{P^2_{\rm eff}D^4},
\end{eqnarray}
and $\epsilon_e=\delta P_e/P_e$,
$\epsilon_{\bar e}=\delta P_{\bar e}/P_{\bar e}$
and $\epsilon_{\cal L}=\delta {\cal L}_{\rm int}/{\cal L}_{\rm int}$
are the relative systematic uncertainties.
\par
One can notice, from Eq.~(\ref{Eq:V-diag}), that systematic
uncertainties in $\sigma_{\rm unpol}$ are induced by
$\epsilon_e$, $\epsilon_{\bar e}$ {\it and} $\epsilon_{\cal L}$, while
those
in $A_{\rm LR}$ and $A_{\rm LR,FB}$ arise from
$\epsilon_e$ and $\epsilon_{\bar e}$ only, and {\it not} from
$\epsilon_{\cal L}$.
Finally, $A_{\rm FB}$ is free from such systematic uncertainties.

Defining the inverse covariance matrix $W^{-1}$ as
\begin{equation}
\left(W^{-1}\right)_{ij}=\sum_{k,l=1}^4 \left(V^{-1}\right)_{kl}
\left(\frac{\partial{\cal O}_k}{\partial\epsilon_i}\right)
\left(\frac{\partial{\cal O}_l}{\partial\epsilon_j}\right),
\end{equation}
with
$\epsilon_i=(\epsilon_{LL},\ \epsilon_{LR},\ \epsilon_{RL},\
\epsilon_{RR})$,
model-independent allowed domains in the four-dimensional CI parameter
space
to 95\% confidence level are obtained from the error contours determined by
the quadratic form in $\epsilon_{\alpha\beta}$ \cite{eadie,Cuypers:1996it}:
\begin{equation}
\label{chi2}
\left(\epsilon_{LL}\ \epsilon_{LR}\ \epsilon_{RL}\ \epsilon_{RR}\right)
W^{-1}
\left( \begin{array}{c}
\epsilon_{LL}\\
\epsilon_{LR}\\
\epsilon_{RL}\\
\epsilon_{RR}
\end{array}\right)=9.49.
\end{equation}
The value 9.49 on the right-hand side of Eq.~(\ref{chi2}) corresponds to
a fit with four free parameters \cite{Groom:2000in,James:1975dr}.

The quadratic form (\ref{chi2}) defines a four-dimensional ellipsoid in the
$\left(\epsilon_{LL},\ \epsilon_{LR},\ \epsilon_{RL},\
\epsilon_{RR}\right)$
parameter space.
The matrix $W$ has the property that the square roots of the individual
diagonal matrix elements, $\sqrt{W_{ii}}$, determine the projection
of the ellipsoid onto the corresponding $i$-parameter axis in the
four-dimensional space, and has the meaning of the bound at 95\% C.L. on
that parameter regardless of the values assumed for the others.
Conversely, $1/\sqrt{\left(W^{-1}\right)_{ii}}$ determines the value
of the intersection of the ellipsoid with the corresponding $i$-parameter
axis, and represents the 95\% C.L. bound on that parameter assuming
all the others to be exactly known. Accordingly, the ellipsoidal surface
constrains, at the 95\% C.L. and model-independently, the range of values
of the CI couplings $\epsilon_{\alpha\beta}$ allowed by the foreseen
experimental uncertainties.

For the chosen input values for integrated luminosity, initial beam
polarization, and corresponding systematic uncertainties, such 
model-independent limits are listed as lower bounds on the mass scales 
$\Lambda_{\alpha\beta}$ in Table~1.
All the numerical results exhibited in Table~1 can be represented
graphically. In Fig.~1 we show the planar ellipses that are 
obtained by projecting onto the six planes 
($\epsilon_{LL},\epsilon_{LR}$), ($\epsilon_{LL},\epsilon_{RL}$), 
($\epsilon_{LL},\epsilon_{RR}$), ($\epsilon_{RR},\epsilon_{LR}$), 
($\epsilon_{RR},\epsilon_{RL}$), ($\epsilon_{LR},\epsilon_{RL}$) 
the 95\% C.L. allowed four-dimensional ellipsoid resulting from 
Eq.~(\ref{chi2}). In these figures, the inner and outer ellipses correspond
to positron polarizations $\vert P_{\bar e}\vert=0.6$ and 
$\vert P_{\bar e}\vert=0.0$, respectively. 
%-----------------------------------------------------------------------------
\begin{table}[ht]
\centering
\caption{
Reach in $\Lambda_{\alpha\beta}$ at 95\% C.L., from the
model-independent analysis performed 
for $e^+e^-\to\mu^+\mu^-$ and $e^+e^-$, 
at $E_{\rm c.m.}=0.5$~TeV, ${\cal L}_{\rm int}=50\,\mbox{fb}^{-1}$ and
$500\,\mbox{fb}^{-1}$, $\vert P_e\vert=0.8$ and 
$\vert P_{\bar e}\vert=0.6$.
}
\medskip
{\renewcommand{\arraystretch}{1.2}
\begin{tabular}{|c|c|c|c|c|c|}
\hline
%&&&&& \\
${\rm process}$ & ${\cal L}_{\rm int}$ & $\Lambda_{\rm LL}$ & 
$\Lambda_{\rm RR}$ &
$\Lambda_{\rm LR}$ & $\Lambda_{\rm RL}$ \\
&$\mbox{fb}^{-1}$& TeV & TeV &TeV  & TeV \\
\hline
\cline{2-6}
 & 50 & $ 35 $ & $ 35 $ & $ 31 $ & $ 31 $
\\
\cline{2-6}
$e^+e^-\to\mu^+\mu^-$
 & 500 &$ 47 $ & $ 49 $ & $ 51 $ & $ 52 $ 
\\ \hline
\cline{2-6}
 & 50 & $ 38 $ & $ 36 $ & $ 54 $ &
\\
\cline{2-6}
$e^+e^-\to e^+ e^-$
 & 500 & $ 51 $ & $ 49 $ & $ 84 $ &  
\\ \hline
\end{tabular}
} % end arraystretch
\label{tab:table-1}
\end{table}
%------------------------------------------------------------------------------
%%%%%%%%%%%%%%%%%%%%%%%%%%%%%%%%%%%%%%%%%%%%%%%%%%%%%%%%%%%%%%%%%%%%%%
\begin{figure}%[htb]
\refstepcounter{figure}
\label{Fig:m}
\addtocounter{figure}{-1}
\begin{center}
\setlength{\unitlength}{1cm}
\begin{picture}(11.5,18)
\put(-1.0,8.0)
{\mbox{\epsfysize=10.0cm\epsffile{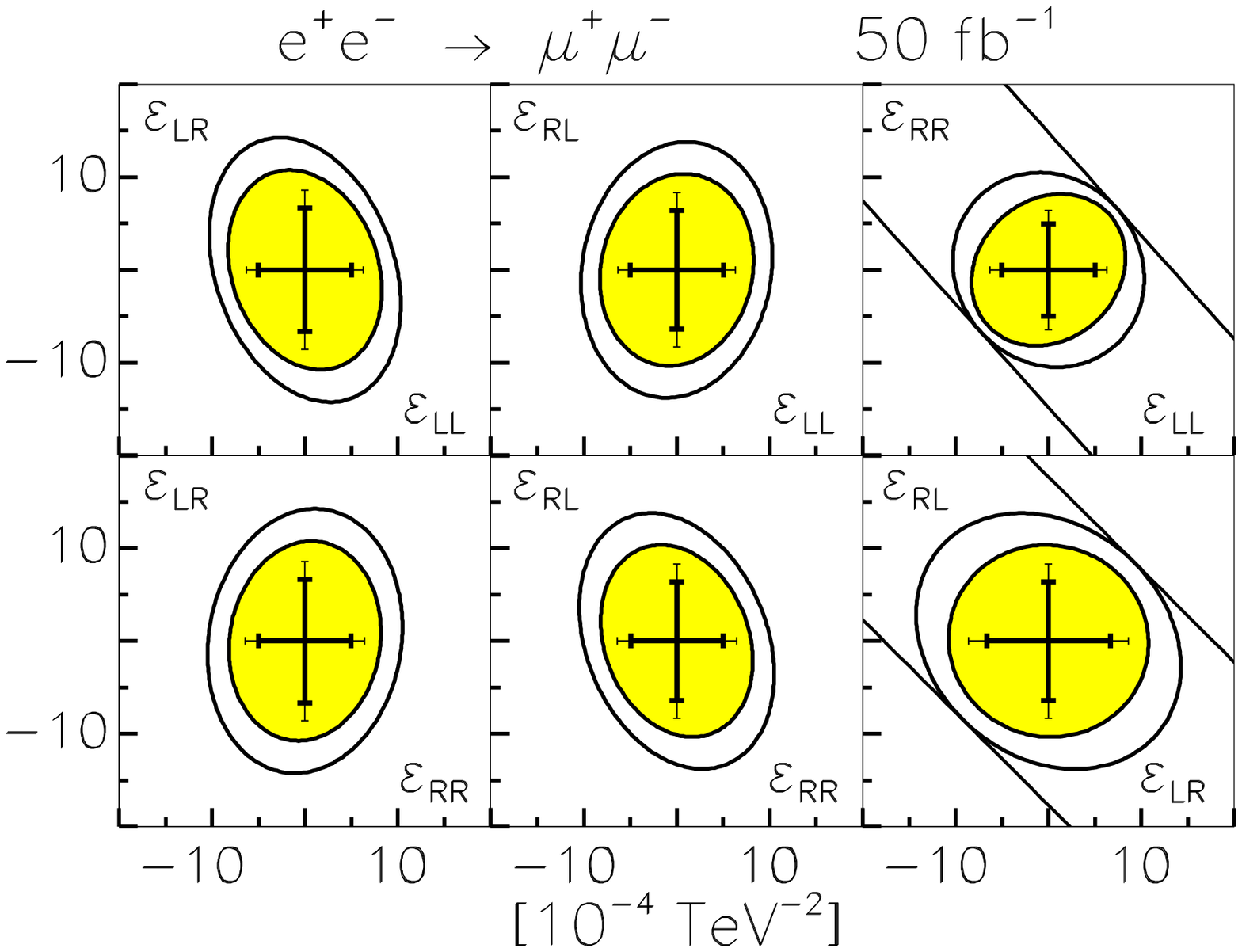}}}
\put(-1.0,-1.0)
{\mbox{\epsfysize=10.0cm\epsffile{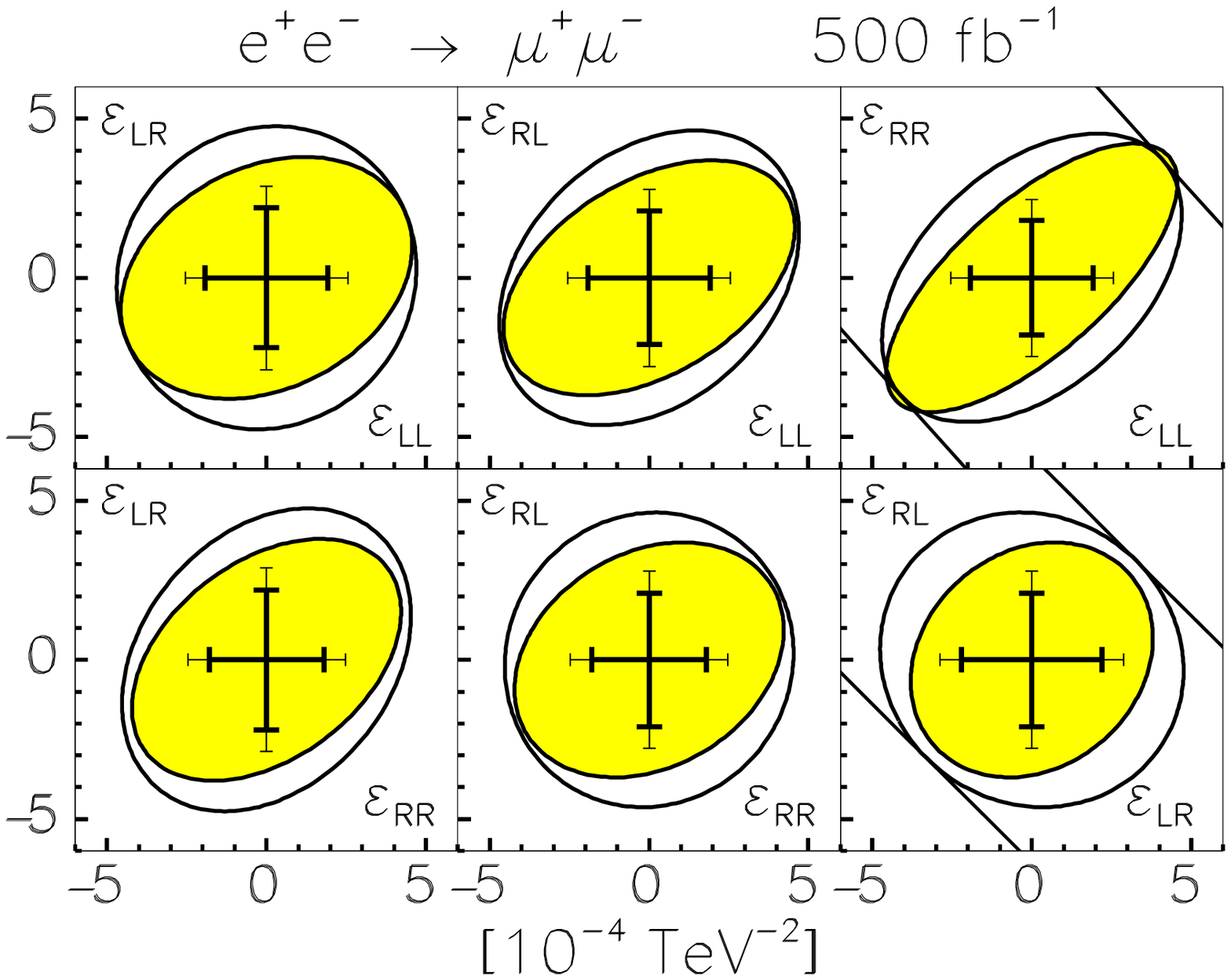}}}
\end{picture}
%\vspace*{-8mm}
\caption{Two-dimensional projections of the 95\% C.L. allowed region 
(\ref{chi2}) for $e^+e^-\to\mu^+\mu^-$ at ${\cal L}_{\rm int}=50~{fb}^{-1}$ 
and ${\cal L}_{\rm int}=500~{fb}^{-1}$.  
$\vert P_e\vert=0.8$, $\vert P_{\bar e}\vert=0.0$ (outer ellipse) and 
$\vert P_{\bar e}\vert=0.6$ (inner ellipse). The solid crosses represent  
the `one-parameter' bounds under the same conditions.}

\end{center}
\end{figure}
%%%%%%%%%%%%%%%%%%%%%%%%%%%%%%%%%%%%%%%%%%%%%%%%%%%%%%%%%%%%%%%%%%%%%%

To appreciate the significant role of initial beam polarization we 
should consider that, in the unpolarized case the only available 
observables would be 
$\sigma\propto (\sigma_{\rm LL}+\sigma_{\rm RR})
+(\sigma_{\rm LR}+\sigma_{\rm RL})$ and 
$\sigma\cdot A_{\rm FB}\propto (\sigma_{\rm LL}+\sigma_{\rm RR})
-(\sigma_{\rm LR}+\sigma_{\rm RL})$, see Eqs.~(\ref{unpol}) and 
(\ref{afbth}). Therefore, by themselves, the pair of experimental 
observables $\sigma_{\rm unpol}$ and $A_{\rm FB}$ are not able to limit 
separately the CI couplings within finite ranges, but could only provide 
a constraint among the {\it linear combinations} of parameters  
$(\epsilon_{\rm LL}+\epsilon_{\rm RR})$ and 
$(\epsilon_{\rm LR}+\epsilon_{\rm RL})$. In some planes, specifically in 
the ($\epsilon_{\rm LL},\epsilon_{\rm RR})$ and 
$(\epsilon_{\rm LR},\epsilon_{\rm RL})$ planes, this constraint has the form
of (unlimited) bands of allowed values, or correlations, such as those 
limited by the straight lines in Fig.~1.
With initial beam polarization, two more physical observables become 
available, {\it i.e.}, $A_{\rm LR}$ and $A_{\rm LR,FB}$, and this enables us
to close the bands into the ellipses in Fig.~1.
The allowed bounds obtained from the observables $\sigma_{\rm unpol}$ and
$A_{\rm FB}$ are not affected by electron polarization (for unpolarized
positrons). Therefore, the bounds in the form of straight lines 
are tangential to the outer ellipses referring to $P_{\bar e}=0$, and in
this case the role of $P_e\ne 0$ is just to close the corresponding 
band to a finite area.
\par 
The crosses in Fig.~1 represent the constraints obtainable
by taking only one non-zero parameter at a time, instead of all four
simultaneously non-zero, and independent, as in the analysis discussed
here.
Similar to the inner and outer ellipses, the shorter and longer 
arms of the crosses refer to positron polarization 
$\vert P_{\bar e}\vert=0.6$ and 0.0, respectively.  Such `one-parameter' 
results are derived from a $\chi^2$ procedure applied to the combination 
of the four physical observables (\ref{unpol})--(\ref{alrfbth}), also 
taking the above-mentioned correlations among observables into 
account. This procedure leads to results numerically consistent
with those presented from essentially the same set of observables 
in Ref.~\cite{Riemann:2001bb},
if applied to the same experimental inputs used there.

\par
For comparison, we also show in Table~1 the corresponding limits
obtained in the case of polarized Bhabha scattering \cite{pp2001}. 
The table shows that for $\Lambda_{\rm LL}$ and $\Lambda_{\rm RR}$ the
restrictions from $e^+e^-\to\mu^+\mu^-$ and $e^+e^-\to e^+e^-$ are 
qualitatively comparable. Instead, the sensitivity to $\Lambda_{\rm LR}$, 
and the corresponding lower bound, is dramatically higher in the 
case of Bhabha scattering. In this regard, this is the consequence of the 
initial beams longitudinal polarization that allows, by 
measuring suitable combinations of polarized cross sections, to directly 
disentangle the coupling $\epsilon_{\rm LR}$. Indeed, without polarization, 
in general only correlations among couplings, rather that finite allowed 
regions, could be derived.

\section*{Acknowledgements}
It is a pleasure to thank Prof. P. Osland and Prof. N. Paver for 
the fruitful and enjoyable collaboration on the topics covered here.
\goodbreak

%%%%%%%%%%%%%%%%%%%%%%%%%%%%%%%%%%%%%%%%%%%%%%%%%%%%%%%%%%%%%%%%%%%%%%

\vspace*{20mm}
\setcounter{footnote}{0}
\end{document}